# Do we need to consider electrons' kinetic effects to properly model a planetary magnetosphere: the case of Mercury

Giovanni Lapenta, KULeuven, University of Leuven, Belgium and Space Science Institute Boulder, USA

David Schriver, Raymond J. Walker, Jean Berchem, Nicole F. Echterling, UCLA, USA

Mostafa El Alaoui, CCMC, NASA and Catholic University of America, USA

Pavel Travnicek, Space Science Laboratory, UC Berkeley USA and Institute of Atmospheric Physics, ASCR, Prague, Czechia

**Abstract**

The magnetosphere of Mercury is studied using an implicit full particle in cell simulation (PIC). We use a hybrid simulation where ions are full particles and electrons are considered as a fluid to start a full PIC simulation where electrons are also particles and follow their distribution function.

This approach allows us to estimate the changes introduced by the electron kinetic physics. We find that the overall macroscopic state of the magnetosphere of Mercury is little affected but several physical processes are significantly modified in the full PIC simulation: the foreshock region is more active with more intense shock reformation, the Kelvin-Helmholtz rippling effects on the nightside magnetopause are sharper, and the magnetotail current sheet becomes thinner than those predicted by the hybrid simulation.   The greatest effect of the electron physics, comes from the processes of particle energization. Both species, not just the electrons, are found to gain more energy when kinetic electron processes are taken into account. The region with the most energetic plasma is found on the dusk side of the tail where magnetic flux ropes are formed due to reconnection. We find that the ion and electron energization is associated with the regions of reconnection and the development of kinetic instabilities caused by counter-streaming electron populations. The resulting electron distributions are highly non Maxwellian, a process that neither MHD nor hybrid models can describe.

## Main Points:

We study the magnetosphere of Mercury using an implicit, three-dimensional, full particle-in-cell simulation.

The currents, interfaces and transition region are made thinner by the electron kinetic effects and become more active.

The greatest effect of including the electron physics is particle energization for both species.

## 3 Introduction

Mercury and Earth are the only inner planets to have strong internal magnetic fields and consequently planetary magnetospheres (Russell et al., 1988a). The overall structure of Mercury's magnetosphere is like Earth's in that a bow shock forms upstream of the magnetopause, cusps form at dayside high latitudes, and an elongated magnetotail with a plasma sheet forms (J. A. Slavin et al., 2008; Zurbuchen et al., 2011). Critically important processes involving particle kinetics such as magnetic reconnection, wave-particle interactions, and non-adiabatic particle motion strongly influence plasma transport, energization, and loss in planetary magnetospheres. Here, we present results from a global kinetic particle-in-cell (PIC) simulation to investigate the dynamics of Mercury's magnetosphere as a coupled, interacting kinetic system with ions and electrons treated self-consistently. We examine the effects of electron kinetics, which can be important for local-global space plasma physics processes (Verscharen et al., 2021), in the context of planetary magnetospheric dynamics.

Mercury has the distinction of having the smallest planetary magnetosphere in our solar system (Kivelson & Russell, 1995; Russell et al., 1988b). Mariner 10 data from the 1970s established that Mercury has an intrinsic magnetic field, and this inference has been confirmed by magnetic field observations from the Mercury Surface, Space Environment, Geochemistry and Ranging (MESSENGER) spacecraft (Anderson et al., 2008). The planet has a dipole moment of 250 nT $R_M^3$ (where $R_M$ is Mercury's radius = 2439 km) and a tilt with respect to the planetary rotation axis of no more than 5º (Anderson et al., 2008, 2011; Jackson & Beard, 1977; Ness et al., 1974, 1975, 1976; Whang, 1977). One of the major findings from the MESSENGER mission is that Mercury's magnetic dipole is offset to the north of the geographic equator by about 0.2 $R_M$ (Anderson et al., 2011). The solar wind interaction with Mercury's intrinsic magnetic field creates a magnetosphere that can divert solar wind plasma around the planet and act as a leaky shield, like Earth's magnetosphere. Due to the relatively weak intrinsic planetary magnetic moment and the highly varying solar wind, the magnetosphere of Mercury is one of the most dynamic in the solar system. This was illustrated during Mariner 10 spacecraft passes through Mercury's magnetosphere that showed extreme changes in the magnetic field and particle bursts occurring on timescales of a few minutes (Ness et al., 1974; Simpson et al., 1974), and more recently by MESSENGER, which showed evidence of rapid and extreme loading and unloading of magnetic flux within Mercury's magnetosphere due to intense magnetic reconnection (J. A. Slavin et al., 2010). Observations have shown that magnetic reconnection occurs constantly at Mercury's dayside magnetopause and nightside magnetotail resulting in flux transfer events, plasmoids and dipolarization fronts (DiBraccio et al., 2013; Imber et al., 2014; J. A. Slavin et al., 2009; James A. Slavin et al., 2012; Sun et al., 2016; Sundberg et al., 2012), representing evidence of substorm-like activity within Mercury's magnetosphere. Multi-scale Kelvin-Helmholtz waves and vortices have been observed along Mercury's magnetopause (Scott A. Boardsen et al., 2010; Gershman et al., 2015), as well as ion Bernstein waves observed in the inner magnetosphere (S. A. Boardsen et al., 2015; Scott A. Boardsen et al., 2012), indicating the importance of kinetic processes in Mercury's magnetosphere.

Since the discovery that Mercury has an intrinsic magnetic field and a dynamic magnetosphere, it has been debated whether Mercury can have an equatorially trapped particle population encircling the planet analogous to Earth's radiation belt (Baker et al., 1987). While Earth's dayside magnetopause subsolar standoff location nominally is 10 – 14 $R_E$, Mercury's is typically 1.35 – 1.55 $R_M$ (Winslow et al., 2013). Using a scaling factor based on the stand-off distances of 8 between the two magnetopause locations, an Earth-like radiation belt or ring current, located at about 4 – 8 $R_E$ radial distance around Earth, would be located at < 1 $R_M$ (inside of Mercury), precluding the existence of a trapped particle population around Mercury. Using simulations of a Mercury-sized magnetosphere and MESSENGER observations (Schriver, Trávníček, Anderson, et al., 2011; Schriver, Trávníček, Ashour-Abdalla, et al., 2011) found that for low to moderate solar wind pressure a relatively low energy (1 – 10 keV), quasi-trapped equatorial population could exist around Mercury at radial distances between about 1.3 – 1.5 $R_M$. This was further supported by observations that revealed the presence of equatorial bulk plasma energies in the 1-10 keV range (Ho et al., 2012, 2016) and more detailed comparison of MESSENGER data and global hybrid simulations (Herčík et al., 2016). Although the bulk of Mercury's inner magnetospheric population have keV energies, higher energy (~100 keV) transient electron bursts were commonly observed on MESSENGER (Ho et al., 2011, 2012, 2016; Lawrence et al., 2015, Baker et al., 2016), although not in the form of a stable, long-lived Van Allen-type radiation belt. The MESSENGER observations clearly show that ion and electron kinetic processes are important at Mercury, but the question remains as to which mechanisms are dominant, and how they shape and influence the structure and dynamics of Mercury's magnetosphere.

Another question that arises out of the observational results is why the overall energy of the quasi-trapped particle belt at Mercury is relatively low (1-10 keV), compared for example to the much higher energies found in Earth's radiation belt (> 100 keV). One possibility is that if energies are too high within the quasi-trapped region at Mercury, the larger ion gyroradius will cause the particles to hit either the planet or the magnetopause and be lost. Another possibility is that due to the overall smaller magnetic field of Mercury, Fermi and betatron acceleration will not be as effective as at Earth (Ashour-Abdalla et al., 2011) due to a smaller ratio between the magnetic field in the trapping region near the planet and the magnetic field in the magnetotail. It is not clear how and where electrons are accelerated in the transient high energy (>100 keV) events. One possible source is the rapid and explosive reconnection that is commonly observed at Mercury. Wave-particle interactions are another mechanism that can lead to overall plasma energization and precipitation loss. The best way to answer these inter-related questions is to use fully kinetic global PIC simulations that include both ion and electron kinetic effects along with MESSENGER data.

Fully self-consistent three-dimensional (3D) global simulation modeling of the interaction of the solar wind with a planetary magnetosphere is challenging due to the vast range in spatial and temporal scales of the physical processes that operate in a plasma. For example, electron kinetic mechanisms scale to the electron plasma frequency and electron Debye length and are the order of seconds and meters, respectively, while ion and large-scale convection processes can occur on planetary scales and take minutes or longer. Global 3D simulations of planetary magnetospheres carried out by using single and multi-fluid magnetohydrodynamic (MHD) models have been successful in modeling the overall magnetospheric structures including the bow shock, magnetopause, cusps and magnetotail (Benna et al., 2010; Dong et al., 2019; Jia et al., 2015; Kabin, 2000). The MHD fluid approximation, however, does not include important ion and electron kinetic effects. Hybrid simulations that treat ions as self-consistent kinetic particles have been used to model the solar wind interaction with magnetized objects and planetary magnetospheres (Aizawa et al., 2021; Fatemi et al., 2017; Kallio & Janhunen, 2003; Karimabadi et al., 2002; Lipatov et al., 2002; Müller et al., 2011; Omidi et al., 2002; P. Trávníček et al., 2007), however, since electrons are treated as a massless fluid, such simulations lack electron kinetic physics. (Aizawa et al., 2021) compared two MHD and two hybrid calculations and found that the models gave the same dayside boundaries (bow shock and magnetopause).

Ongoing rapid advances in computational speed and memory using massively parallel processing supercomputers combined with improved numerical techniques such as implicit PIC codes, now allow realistic fully self-consistent ion-electron global 3D PIC simulations of small planetary magnetospheres. In this paper, we use the iPic3D plasma simulation code (Lapenta, 2012; Markidis et al., 2010a; Vapirev et al., 2013) to study the evolution of Mercury's magnetosphere. It is a fully kinetic, electromagnetic implementation of the implicit moment approach that allows the description of macroscopic scales while resolving the electron skin depth and gyroradius (Brackbill & Forslund, 1982; Lapenta, 2012; Lapenta et al., 2006). The advantage of implicit PIC codes is that the electron plasma frequency and Debye length do not need to be resolved, but electron kinetics are still included. Comparing results from the global iPic3D code which includes both electron and ion kinetics with existing global hybrid simulations of Mercury's magnetosphere, provides a unique way of separating out the effects of electrons on global magnetospheric dynamics and investigating electron and ion energization.

## 2 Approach

Our approach is to start an implicit full PIC simulation using a quasi-steady state reached at the end of a global hybrid simulation run fully described in (P. M. Trávníček et al., 2010). As reported there, this state is not subject to rapid changes and provides a good starting point for our investigation of the effects added by the electron kinetic physics. This reference simulation used the global hybrid code (*Travnicek et al., 2009, 2010; Herčík et al., 2013, 2016)*. We use a 3-D simulation box composed of 940×400×400 mesh points distributed along the three (Cartesian) dimensions with the spatial resolution $\Delta x=0.4 c/\omega_{ppsw}$, $\Delta y=\Delta z=c/\omega_{ppsw} \equiv d_{psw}$, i.e., the size of the simulation domain is $375 \times 400 \times 400 d_{psw}^3$. Here $c$ is the speed of light, $\omega_{ppsw}$ and $d_{psw}$ are the solar wind proton plasma frequency and proton inertial length, respectively. The $X+$ axis is oriented along the solar wind flow and the axis $Z-$ is parallel with the magnetic moment vector of the Mercury. The $Y+$ axis completes a right-handed coordinate system. Macro-particles are advanced with the time step $\Delta t=0.02 \Omega_{psw}^{-1}$ (where $\Omega_{psw}$ is the solar wind proton gyrofrequency) whereas the electromagnetic fields are advanced with $\Delta t_B = \Delta t/10$.

We initialize the magnetic field with a superposition of the interplanetary magnetic field (IMF) and a dipolar planetary magnetic field. The IMF is $B_{sw}=(B_x,0,B_z)=(0.939,0,0.342)$, where $B_{sw}=1$, makes an angle $\phi=+20°$ with respect to the $+X$ axis (i.e., with respect to the solar-wind flow direction) and points toward the planet and north. The dipolar field is by $B_M = (M/r^3)[-2\sin\lambda\, e_r + \cos\lambda\, e_\lambda]$, where the magnetic moment $M$ and radial distance $r$ from the center of Mercury are $B_{sw} d_{psw}^3/\mu_0$ and $d_{psw}$, respectively, $e_r$ and $e_\lambda$ are unit vectors in the radial and magnetic latitude directions, respectively, and $\lambda$ is the magnetic latitude measured from the equatorial plane $(X,Y)$ (no tilt of the planetary dipole is applied, however, the dipolar magnetic field is shifted in the north pole direction by $0.2 R_M$). We use a scaled-down model of Mercury with a magnetic moment $M=100{,}000\, B_{sw} d_{psw}^3 4\pi/\mu_0$.

The downscaling compromises the ratio between particle gyroradius and the radius of Mercury. However, because Mercury has a the relatively big radius and is immersed into a relatively high-density solar wind this fact does not compromise the physics qualitatively. For example, the standoff distance of Mercury's bow shock from the surface of the model Mercury is unchanged in the downscaled model. Note, that the typical size of Mercury varies between $60-90 d_{p,sw}$ according to the value of the solar wind density $n_{p,sw}$ thanks to the higher eccentricity of its orbit around Sun (and its close proximity to the Sun).

At $t=0$, we load the simulation box with macro-particles in each cell (except for the interior of the planet) representing a Maxwellian isotropic proton plasma of density $n_p=n_{psw}=15$ with bulk speed $v_p=(v_{sw},0,0)$, where the solar wind speed $v_{sw}=4\,v_{Asw}$ ($v_{Asw}$ is the Alfvén speed in the solar wind). The ratio of kinetic to magnetic pressure $\beta_{psw}=2\,n_{psw}T_{psw}/B_{sw}^2=0.5$, where $T_{psw}$ is the solar wind plasma temperature. Electron beta is $\beta_e=1.0$. This plasma flow is continuously injected from the left boundary into the simulation box.

In the reference hybrid run used in the PIC simulation (Trávníček et al., 2010), , the solar wind speed arriving at Mercury is 450 km/s, the solar wind density is 15 cm³, the total $B$ for the IMF is 18nT and the angle between *x* and *z* is 20 degrees with $B_x=16.7nT$ (toward Mercury), $B_z=6.7nT$ (northward) and $B_y=0$. Starting from the hybrid run state, we run a PIC simulation spawned from it. Instead of reusing the ions from the hybrid run, we chose to generate the particles from the moments of the hybrid run. We set the densities for both species to be equal to their values in the hybrid simulation thereby assuming quasi-neutrality. The ion fluid velocity is directly available and the electron velocity relative to the ions can be determined from the current needed to support the ∇×**B** term from Ampere's law. We assume that the initial electron temperature was identical to the ion temperature from the hybrid simulation. This provides three different velocity values for the parallel and the two perpendicular directions. In addition to providing the initial state of the fields and the particles, the hybrid simulation also provides the boundary conditions at the planet and at the solar wind inflow. At the other boundaries, we use open boundary conditions obtained from Richardson extrapolation of the inner field (Wan & Lapenta, 2008). This approach is a direct generalization of our previous multi-scale approach to generate a PIC run from a MHD state (Walker et al., 2019) but uses the additional information on the anisotropic ion temperature. The boundary condition around the planet (represented by the spherical green shell in the figures is) is applied by using the fields and moments from the hybrid reference state.

Global PIC simulations of the interaction of the solar wind with a planetary magnetosphere are challenging. Approximations have to be made. Although we consider a domain of 9.6 x 12 x12 $R_M$, we have to re-scale it to make the simulation feasible with a PIC approach that fully resolves the kinetic physics of electrons and ions. The ratio between the ion skin depth and $R_M$ is reduced by a factor of 10 to 3.3, therefore Mercury's radius is 6.6 times the ion skin depth. This assumption can appear to be a limitation but previous literature (Tóth et al., 2017) has shown that provided the scales remain well separated, the physics is qualitatively correct, if not quantitatively precise. Increasing the grid points 10-fold in each direction would increase computing resources by a factor of 1000, which is not yet feasible. Alternatively, we could increase the grid spacing but then the electron physics would not be resolved. Another accepted practice to carry out large scale computations is to reduce the mass ratio of ions to electrons ($m_i/m_e$). However, altering the mass ratio would also alter the ratio between ram and thermal pressure of the incoming solar wind and thus affect significantly the results of the simulation. We avoid this problem by using the exact physical mass ratio ($m_i/m_e = 1836$).

The full PIC simulation box is divided into 160x200x200 cells and uses $7^3$ (343) particles per cell per species. The time step is $dt=2.4 \cdot 10^{-3}$s. The total simulation lasts 1600 cycles and covers 3.8s of real time. As the results shown below indicate, this is sufficient time for the electron state to change from the fluid state in the hybrid simulation to a kinetic state. The fastest mean electron speed is sufficient to cross the simulation box in the *x*-direction twice, i.e. two electron crossing times. We used the iPic3D code (Markidis et al., 2010b) with the Ecsim energy conserving implementation (Gonzalez-Herrero et al., 2018; Lapenta, 2017) and the correction to impose charge conservation (Chen & Tóth, 2019). A decentering time scheme of $\Theta=0.5$ ensures exact energy conservation (Lapenta, 2017). This feature is critical to correctly capture particle energization. By comparison, a run with $\Theta=1.0$ severely looses energy and leads to only feeble particle energization: energy conservation is key to obtaining correct results in this case. The run required 12 hours of wall-clock time on 3200 processors. While this is very far from real-time, it is a relatively manageable computational effort. The coordinate system has *x* positive in the direction away from the Sun, *y* positive toward dawn and *z* positive north. It's worth recalling that unlike Earth, the dipole of Mercury is shifted vertically very significantly, leading to a significant north-south asymmetry.

Figure 1 shows a three dimensional overview of the state of Mercury's magnetosphere at the end of the the fully kinetic simulation (b) compared with the state from the hybrid simulation used for its initialization (a). A selection of magnetic field lines is shown along with a cross section in the YZ plane at $\frac{x}{R_H} = 6.45$ showing the net current (sum of electron and ion current). As can be seen, the overall structure of the magnetosphere is similar in the two simulations.

The bow shock and foreshock regions are visible. The pronounced north-south asymmetry is a well-known property of Mercury's magnetosphere due to the displacement of the magnetic dipole of the planet away from its center. For the solar wind configuration used in this study, the southern magnetosheath and foreshock region are much more dynamic than in the north.

## 3 Impact of the electrons on macroscopic scales

One of the motivations of our study was to determine whether adding electron kinetic physics introduces significant changes to the global features predicted by the hybrid simulation. We analyzed the overall state and found that the changes introduced by the electrons are only local.

The most notable difference visible in Figure 1 is the increased current in the full kinetic simulation: the color scale for the full PIC run is 10 times larger than in the hybrid state. The presence of the electrons leads to a more localized current at the magnetopause and in the tail. This effect is well known and studied for the Earth (*Berchem and Okuda, 1990*).. The interaction between the heliospheric plasma and the magnetic field of the planet leads to a diamagnetic effect with the separation of the incoming plasma from a relative vacuum formed by the planetary magnetosphere. The thickness of this transition at the magnetopause is determined by a complex interplay of electron and ion scales (*Berchem and Okuda, 1990*). In hybrid models, the thickness is determined by the ions, but in presence of the electrons the interface becomes thinner with the physics of the electron gyromotion playing a crucial role. A recent study (*Park et al., 2019*) noted that as the mass ratio is changed artificially from the physical value of 1836 down to the case of an electron positron plasma, the thickness remains determined by the electron gyro scale. This effect is completely missed by hybrid models.

Figure 2 reports a synthetic fly through at the end of the full kinetic simulation along the vertical white line identified in Figure 1. As can be seen the tail current is almost entirely carried by the electrons. At the magnetopause, the electron current causes an intensification and a narrowing of the layer. The role of the electrons does not modify the global structure but introduces important local effects.

This can be seen in Figures 3-5 where we compare maps of the three components of the magnetic field from the hybrid simulation to those at the end of the PIC simulation. The overall magnetic structure remains the same. The locations of the major boundaries (bow shock, magnetopause) and the configuration of the magnetotail are unchanged, but some interesting differences brought by the electron physics appear when examining the results in detail.

The foreshock region sunward of the planet is somewhat more active with more intense shock reformation. Although shock reformation is present in the hybrid case, it is more visible in the structures forming ahead of the shock in the kinetic results. The magnetopause is less affected. The Kelvin-Helmholtz ripples in the night side part of the magnetopause (most evident on the dusk side of the XY plots in Figures 3 and 5) are made sharper by the electrons. This is a direct consequence of the thinner and more intense current layer formed by the electrons, shown in Fig 1-2. The growth rate of a Kelvin-Helmholtz instability increases as the layer becomes thinner (*Chandrasekhar, 1961*).

The current sheet in the tail becomes thinner as is to be expected because of the more intense current brought by the electrons (see the XZ plane in Figure 3). The dawn-dusk field is the smallest of the three components (Figure 4) and for this reason is more affected by the electrons, especially in the night-side region. The dawn-dusk field spreads out more widely in the lobe regions, especially in the northern hemisphere (see Figure 4 YZ plots).

Another opportunity afforded by the full kinetic model is a more complete description of reconnection. Figure 6 shows a blow up of the tail region at the end of the full kinetic run. Shown are selected magnetic field lines superimposed on a XY cross section of the false color representation of the electron agyrotropy (*Scudder and Daughton, 2008*). We show here one of the regions of tail reconnection but all other regions of intense agyrotropy are associated with shearing of magnetic field lines possibly related with 3D complex reconnection processes or component reconnection. Agyrotropy is a property of the electron pressure tensor and is related to the motion of the electrons in the plane normal to the local magnetic field. When agyrotropy is large, the motion is not one of simple gyration leading to a non-isotropic distribution in the perpendicular plane. Hybrid models based on fluid electrons cannot capture this process that is key to the kinetic reconnection process (*Biskamp, 1996*; *Burch et al, 2016*).

Agyrotropy gives a good marker to identify the main interfaces of the magnetosphere because it is sensitive to the localized changes in the velocity distribution due to finite Larmor effects. In the dayside, we can clearly identify the bow shock. For Mercury, the vicinity of the planet with the bow shock makes the region of the bow shock, magnetosheath and magnetopause appear as a bright region of yellow in Fig. 6, where the electrons are non gyrotropic. In the nightside the magnetopause is sharply identified by agyrotropy. The region inside the nightside magnetopause is the Hermian tail where we focus our study of reconnection. Figure 6 focuses on one specific reconnection site on the dusk side, the same we will concentrate on later to study energy exchanges. The topology of this reconnection site is shown by the field lines. Some field lines are already reconnected and connected to the planet; others are also reconnected but are instead linked to the tailward end. In between we observe lines being reconnected and forming a flux rope in the dawn-dusk direction. The color of the field lines is based on the value of the $B_x$ component facilitating following the orientation of the field along the lines. This is a typical case but other regions in the tail characterized by strong agyrotropy present similarly complex field lines that indicate reconnection. The details of the topology of the field lines in different regions, however, differ greatly in detail presenting an interesting opportunity for additional future extensive analysis of all possible reconnection site, using for example automatic techniques (Parnell et al., 2008; Olshevsky et al., 2016; Lapenta, 2021; Sisti et al., 2021). We focus here on more global energy transfers; future work will investigate the topology of reconnection in more details.

On a global scale, three things can be learned from this simulation. First, electron scale physics does not introduce radical modifications to the global configuration of the magnetosphere, leaving unchanged the locations of the dissipative boundaries and the overall field configuration. The effects of the electrons are mostly localized at the interfaces: bow shock, magnetopause and in the magnetotail current sheet. Second, the fluid treatment of the electrons used by hybrid models provides an effective way to model the electron contribution to the global interaction of the solar wind with planetary magnetospheres but it fails to provide the details of the localized features and in particular the physics of reconnection. Finally, scaling down the planet size with respect to the ion skin depth which makes the fully kinetic description possible, did not damage the quantitative accuracy of the kinetic model: it still captures the global scales correctly.

## 4  Effect on particle energization

Another goal of our study was to determine the effects that electrons have on processes at both ion and electron scales. We found that the changes in particle energization are significant.

Our results show that only a fully kinetic model can provide reliable information on particle energization. Of course, the effects are stronger on the electrons, but the ions also are significantly energized by electron effects, especially in the magnetotail region. The contribution of the electron physics to the energization of both species is especially perceivable near the boundaries (e.g., bow shock, magnetopause) and in the magnetotail. Figure 7 shows the thermal energy of the two species defined as

$$E_{th,s}(\mathbf{x},t) = \frac{m_s}{2} \frac{\int (v - \langle v \rangle)^2 f_s(\mathbf{x},\mathbf{v},t)d\mathbf{v}}{\int f_s(\mathbf{x},\mathbf{v},t)d\mathbf{v}} \quad (1)$$

where $m_s$ is the mass of particle species $s$, electrons and protons in our case. In (1) $f_s$ is the distribution function for species $s$, $v$ is the velocity and $\langle v \rangle$ is the bulk velocity. This quantity has the units of energy. Figure 7 gives the thermal energy (keV) in the XY (equatorial) plane. Both species gain considerable energy compared with the hybrid case. On the dayside, the energy gain occurs at the bow shock and is localized downstream of it and upstream of the magnetopause. In the tail, the energy gain starts in the vicinity of the planet and extends into the tail on the dusk side where reconnection occurs.

In Figure 8 we show the energy for the two species along the equatorial and noon-midnight planes. Plotted are the thermal energy for electrons (left), the thermal energy for ions (center) and the total energy for ions.   For the electrons, the thermal energy is dominant and the bulk energy is negligible. For the ions, the bulk energy is very substantial, exceeding the thermal energy as can be seen by examining the difference between the total and thermal energies.   The ram part of the ion energy is dominant as the solar wind is relatively cold. The region with the most energetic plasma is on the dusk side of the tail. In the magnetosheath, the ion energy is relatively high on both the dusk and dusk sides but larger on the dusk side. In particular, the dusk side of the tail shows the formation of magnetic islands due to reconnection. The ion and electron energization are associated closely with the regions of reconnection. The difference in energization of the species in a full kinetic model leads to an additional insight: electrons are more significant in the transfer of energy within the system.

Figure 9 shows the $x$ directed (away from the Sun) energy flux. The energy flux is defined as:

$$\mathbf{Q}_s(\mathbf{x},t) = \frac{m_s}{2} \int vv^2 f_s(\mathbf{x},\mathbf{v},t)d\mathbf{v} \quad (2)$$

In the hybrid run (top row), all energy flux is carried by the ions while the electron energy flux is negligible. In the full kinetic run, the ion energy flux is left largely unchanged (middle row) but the electrons now also carry a significant energy flux (bottom row). The electron energy flux is especially strong in the magnetosheath and the near-planet tail. As we show in the next section, the electron energy flux is linked with two unequal counter streaming beams that lead to a net energy flux.

## 5 Mechanism of energization in the full kinetic model

The key modification introduced by the electron physics is the energization of the electrons and to a lesser degree the ions, in the dayside magnetosheath and in the magnetotail region. In this section, we discuss the cause of the electron energization by studying the electron velocity distributions in these two regions. In Figure 10 and Figure 11, we plot the electron distributions in magnetic coordinates, respectively for the dayside and the magnetotail region. We select the region of most evident reconnection, observed in Fig. 7 and 8 near the position $y/R_H=7.5$ and $z/R_H=7.5$. Two positions along x are chosen, one for the tail and one for the dayside and are indicated in Fig. 11 and 12.

On the dayside, the distribution remains gyrotropic as evidenced by the circular symmetry in the perpendicular plane. The acceleration occurs in the parallel direction.

On the night side, two populations are evident: one traveling in the direction of the magnetic field and one traveling in the opposite direction. Recalling that these particles are line tied to magnetic field lines that map to the polar regions on the planet, the acceleration mechanism is linked with the interaction between the particles travelling towards the planet and those reflected back via magnetic mirroring. The particles travel back and forth on the same field lines forming two interacting counter-streaming populations. This leads to the two-stream instability that tends to fill the velocity space gap between the two populations forming flattened ellipsoidal distributions. A similar process is occurs in   Earth's magnetotail for the ions reflected in dipolarization fronts (*Birn et al, 2017)* and entering from the northward and southward separatrices (*Aunai et al, 2011; Lapenta et al., 2016*). The same process is also present on the dayside where the parallel distribution is much wider than the perpendicular distribution, pointing again to parallel acceleration and counter-streaming beams.

Streaming instabilities cause the redistribution of the velocity between the two counter-streaming beams via the action of the electric fields generated by the instability. To represent them correctly, a fully kinetic description of the electrons through the entire domain is needed. Our study shows that it is not sufficient to just consider a sub-domain of interest. Previous studies have concentrated on the tail region or the dayside region alone, embedding a kinetic model of a portion of the magnetosphere from a global model. Such an approach cannot be used here because the mirroring process need to be included: the kinetic box must be resupplied with electrons distributed in a non-Maxwellian two-beam distribution.

# 6 Conclusions

We studied the magnetosphere of Mercury using a full kinetic approach based on the code iPic3D. The simulation is initialized starting from a state obtained from a global hybrid simulation. The modification of the hybrid state caused by the full kinetic description of the electrons is followed for 3.8s, a time that exceeds by many orders of magnitude the electron kinetic scales, allowing the electron kinetic physics to fully develop.

The overall global evolution is not significantly affected by the presence of kinetic scales. The electron scale physics does not change radically the global configuration of the magnetosphere, and the main features of it remain similar: the interface locations and the overall field configuration remain the same on the time scale considered in this study.

The localized features of the magnetospheric boundaries are modified substantially. The bow shock, the magnetopause and the magnetotail current sheet become sharper and thinner. Their sharper nature leads to an increase in the wave activity occurring in their proximity: the foreshock becomes more active and the Kelvin-Helmholtz ripples in the night side part of the magnetopause are made sharper by the electron physics.

The greatest change introduced by the kinetic physics is in the energization of both species, electrons and ions. Surprisingly, the effect is not only on the electrons but extends also to the ions. Both ions and electrons gain considerable energy compared with the hybrid case. On the dayside, the energization is significantly increased downstream of the bow shock and upstream of the magnetopause. In the night side, the increased energization is localized closer to the planet but extends tailward on the dusk side where reconnection occurs.

The processes of energization are found to be linked with the occurrence of magnetic reconnection and with the presence of populations of counter-streaming electrons. We note that this process can only be correctly included if electron populations are allowed to mirror naturally at the vicinity of the planet. The full kinetic model shows that the resulting electron distributions are highly non Maxwellian, a process that neither MHD nor hybrid models can describe.

While the results presented here are relative to Mercury's magnetosphere, the processes identified are likely to hold true for larger planets like Earth. The rapid evolution of computing power is making it progressively more feasible to extend the study presented here to our home planet.

**Acknowledgments**

This research received funds from the NASA grants HSR 80NSSC19K0841 and HSR 80NSSC19K0846 and NASA SSW 80NSSC19K0789 and 80NSSC21K0053.

Resources supporting this work were provided by the NASA High-End Computing (HEC) Program through the NASA Advanced Supercomputing (NAS) Division at Ames Research Center and NCCS at NASA Goddard Space Flight Center

## 7   Figure Captions

Figure 1: Overview of the state of the Hermian magnetosphere from the reference hybrid simulation (panel a) and from the fully kinetic simulation (panel b), taken at the final cycle 1600 (or 3.8s after the start). The Sun is to the right. The bow shock and foreshock are visible to the right while the tail with its reconnection-generated flux rope is visible to the left. Mercury's inner boundary is shown as a green sphere. The color code gives the Bx component of the magnetic field.

Figure 2: Synthetic crossing along the white line shown in Fig. 1 at the same time. From top to bottom reported are a) the amplitudes of the ion, electron and total currents, b) the three components of the magnetic field, c-e) the x, y and z components of the ion, electron and total currents.

Figure 3: Comparison of the $x$ component of the magnetic field, $B_x$ from the hybrid state used to spawn the kinetic run (top) and the final time of the fully kinetic simulation at cycle 1600 (or 3.8s after the start) (bottom). The lines are projections of magnetic field lines. From left to right three planes are shown (XY) at $z=L_z/2$, (XZ) at $y=L_y/2$ and (YZ) at $x=6R_H$. Note that the spatial scale for the YZ plane ($x=6R_H$) is not the same as in the XY and XZ cuts.

Figure 4: Comparison of the $y$ component of the magnetic field, $B_y$ from the hybrid (top) and fully kinetic (bottom). The display is the same as in Figure 2

Figure 5: Comparison of the $z$ component of the magnetic field, $B_z$ from the hybrid (top) and full kinetic (bottom). The display is the same as in Figure 2.

Figure 6: Blow up of the region of tail reconnection. Magnetic field lines undergoing reconnection are superimposed to a false color representation of the electron agyrotropy on the XY plane (at $z/R_H= 6.8$). The filed lines shown are emanating from a region of radius $r/R_H =0.5$ around $x/R_H =6.8$, $y/R_H =7.2$ and $z/R_H =6.6$.

Figure 7: Evolution of the thermal energies, $E_{th,s}$, in the XY plane from the hybrid (top) and 2 subsequent times from the fully kinetic calculation: second row, $Cycle$=100, third row, $Cycle$=1000. The left (right) column shows the electrons (ions). The white lines magnetic field lines projected on the XY plane.

Figure 8: Species energy at the final time (at cycle 1600 or 3.8s after the start) for electrons (left) and ions. The top row shows the *XY* plane and the bottom the *XZ* plane. From left to right shown are $E_{th,e}$, $E_{th,i}$ and $E_i$. The total energy of the electrons is identical and is not reported because the electron bulk energy is negligible. The white lines are projected field lines.

Figure 9: Energy fluxes in the *x* direction $Q_x$. The top row reports the ion energy flow $Q_{x,i}$ from the hybrid run. The next two rows show the final (*Cycle*=1600) energy flow from the full kinetic simulation, for ions $Q_{x,i}$ (middle row) and electrons $Q_{x,e}$ (bottom row).

Figure 10: Tail. Reduced electron velocity distribution functions at the location *x* shown in lower right panl by an orange line and $y/R_H=7.5\pm Ly/120$ and $z/R_H=7.5\pm Lz/120$. Three different reduced distributions are shown in magnetic coordinates. Note that compared with Figure 11 the scale of the axes have been expanded to account for the higher thermal speed. The plot of the three components of the magnetic field and the magnitude is along the sun-Mercury line.

Figure 11: Dayside. Reduced electron velocity distribution functions at the location *x* shown in the lower right panel with an orange line and $y/R_H=7.5\pm Ly/120$ and $z/R_H=7.5\pm Lz/120$. Three different reduced distributions are shown in magnetic coordinates. The lower right plot shows the three components of the magnetic field and the magnitude along the sun-Mercury line.

# 8 Figures

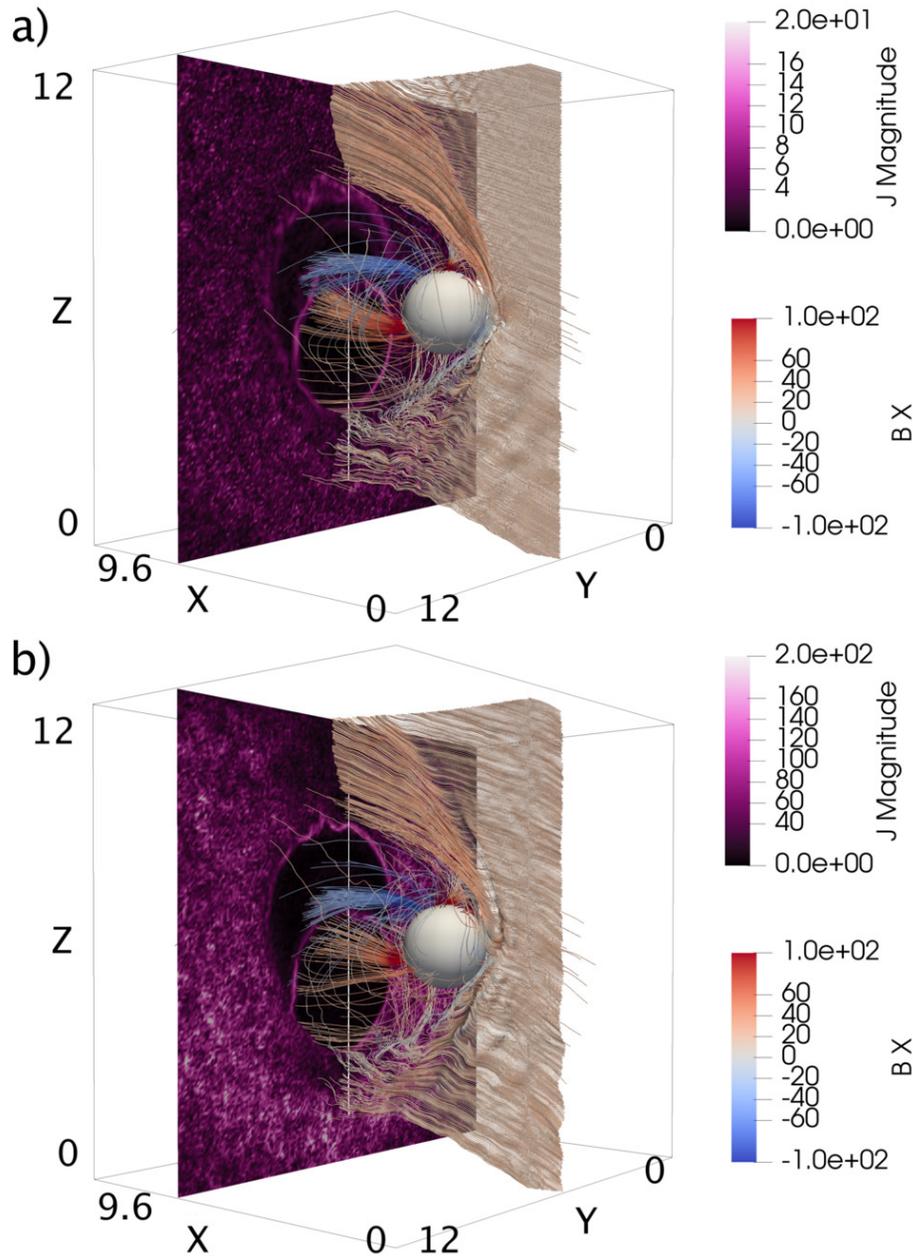

Figure 1: Overview of the state of the Hermian magnetosphere from the reference hybrid simulation (panel a) and from the fully kinetic simulation (panel b), taken at the final cycle 1600 (or 3.8s after the start). The Sun is to the right. The bow shock and foreshock are visible to the right while the tail with its reconnection-generated flux rope is visible to the left. Mercury's inner boundary is shown as a green sphere. The color code gives the Bx component of the magnetic field.

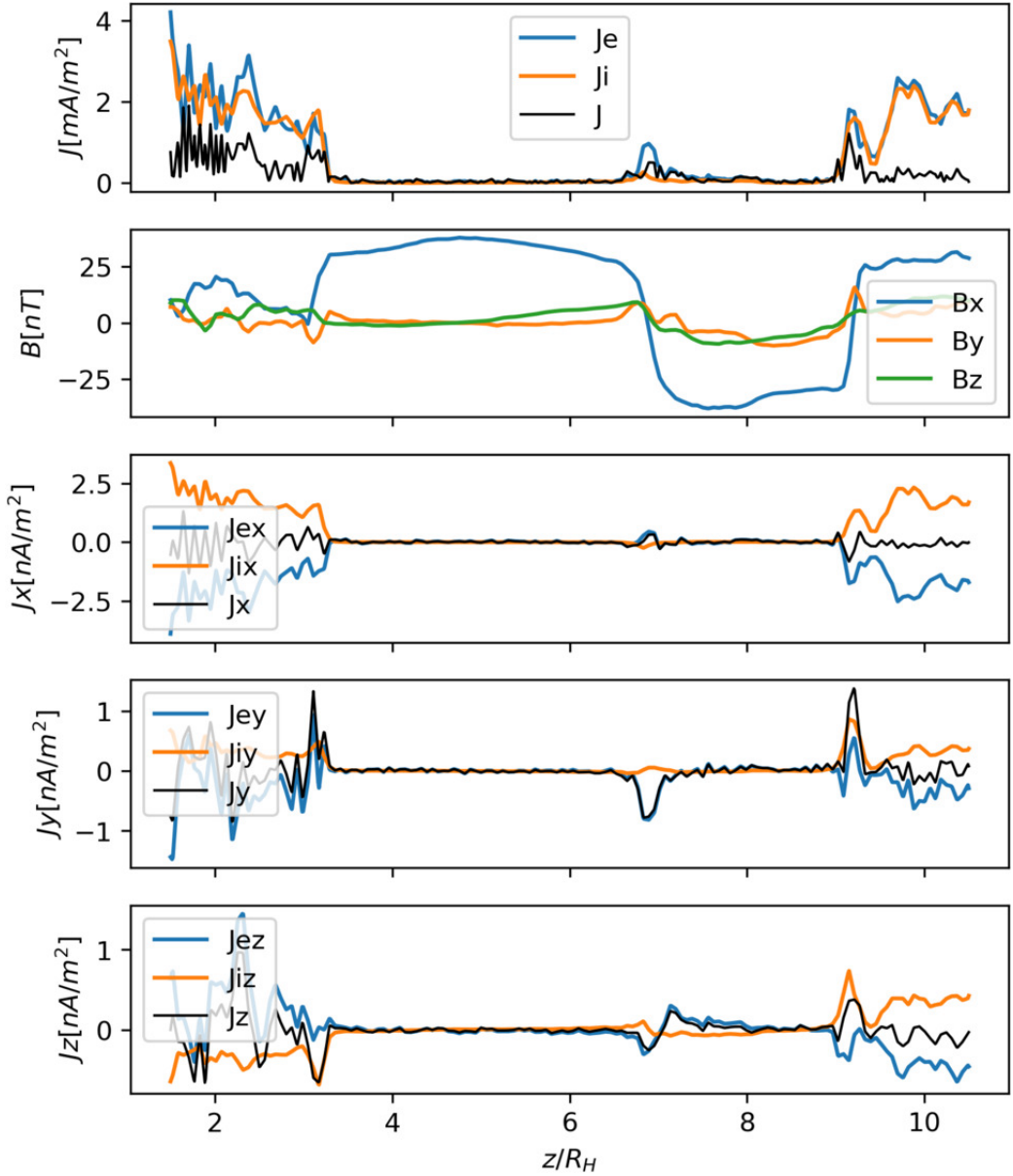

Figure 2: Synthetic crossing along the white line shown in Fig. 1 at the same time. From top to bottom reported are a) the amplitudes of the ion, electron and total currents, b) the three components of the magnetic field, c-e) the x, y and z components of the ion, electron and total currents.

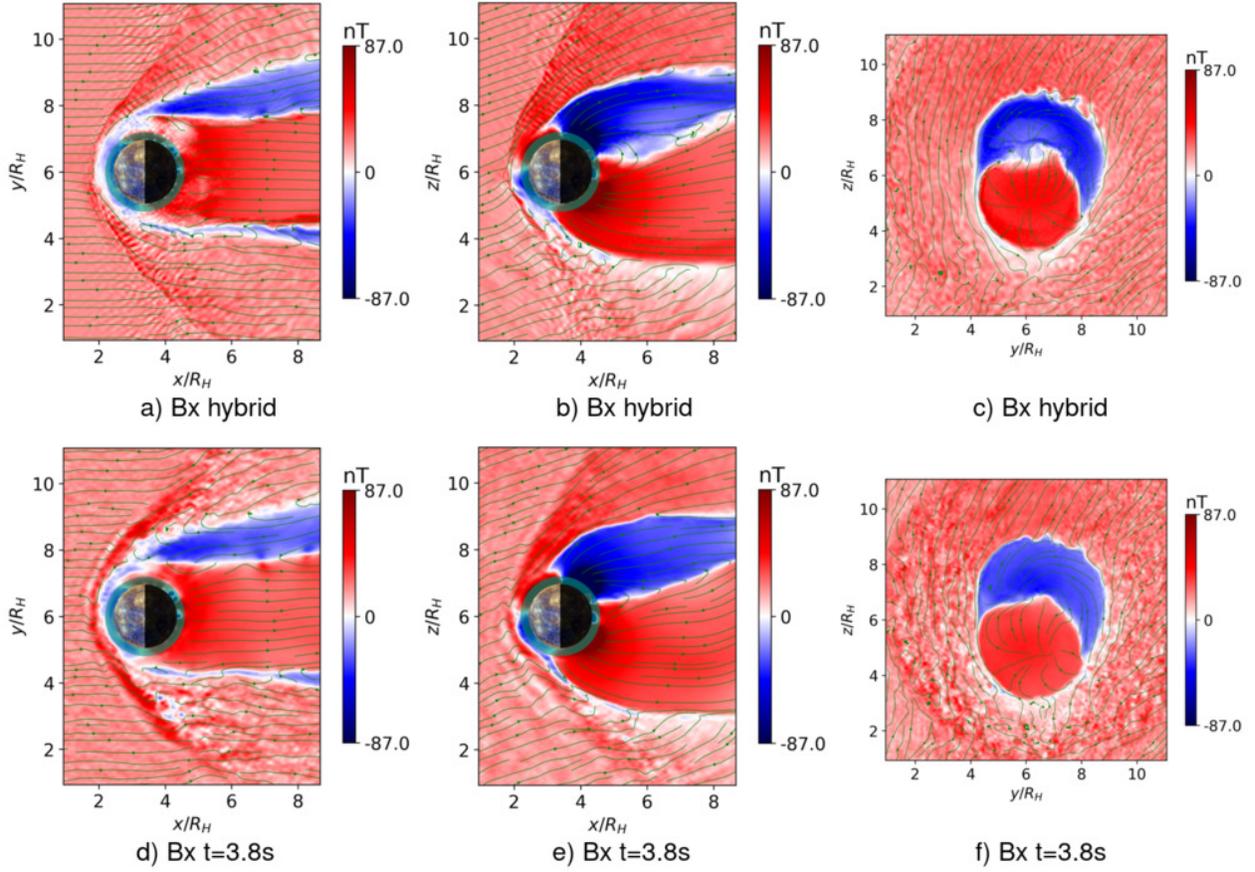

Figure 3: Comparison of the *x* component of the magnetic field, $B_x$ from the hybrid state used to spawn the kinetic run (top) and the final time of the fully kinetic simulation at cycle 1600 (or 3.8s after the start) (bottom). The lines are projections of magnetic field lines. From left to right three planes are shown (*XY*) at $z=L_z/2$, (*XZ*) at $y=L_y/2$ and (*YZ*) at $x=6R_H$. Note that the spatial scale for the YZ plane ($x=6R_H$) is not the same as in the XY and XZ cuts.

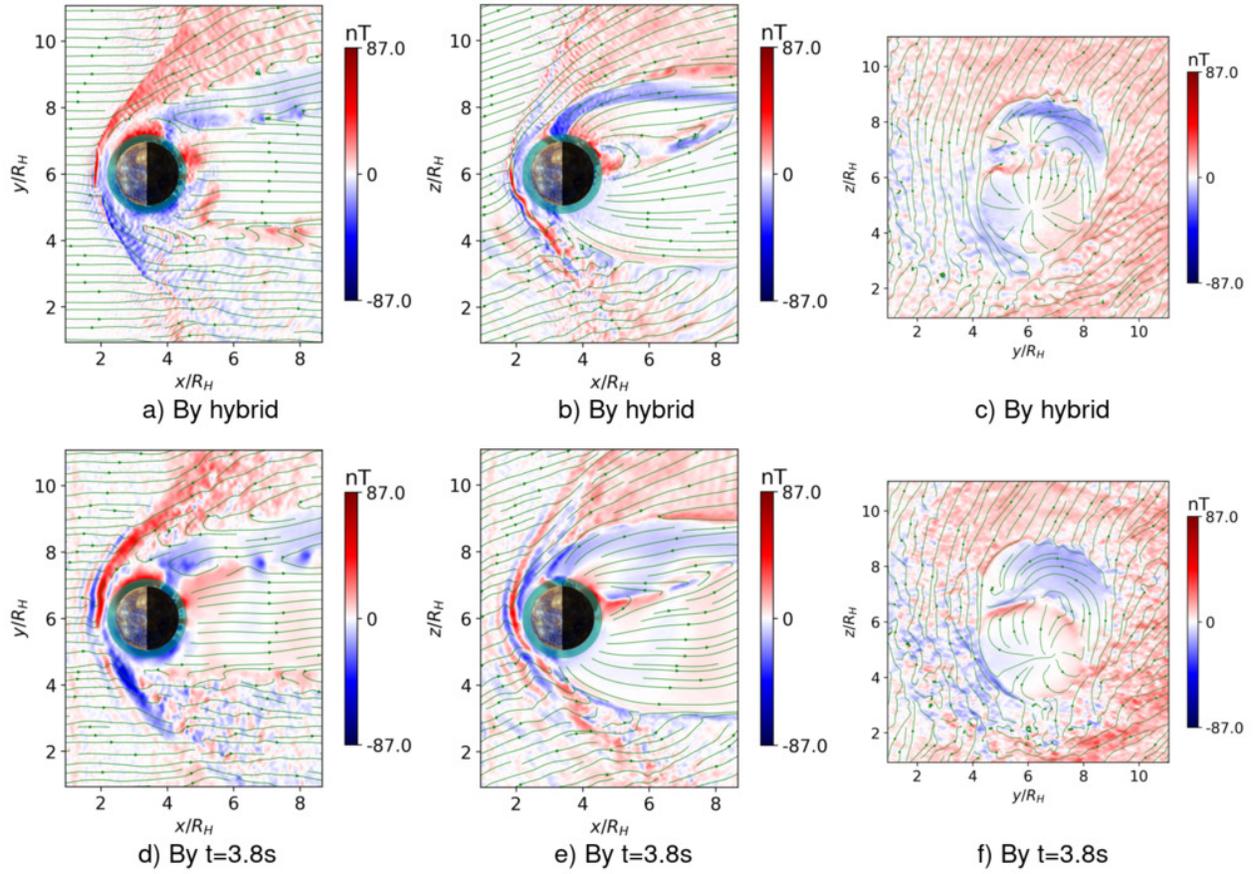

Figure 4: Comparison of the *y* component of the magnetic field, $B_y$ from the hybrid (top) and fully kinetic (bottom). The display is the same as in Figure 2

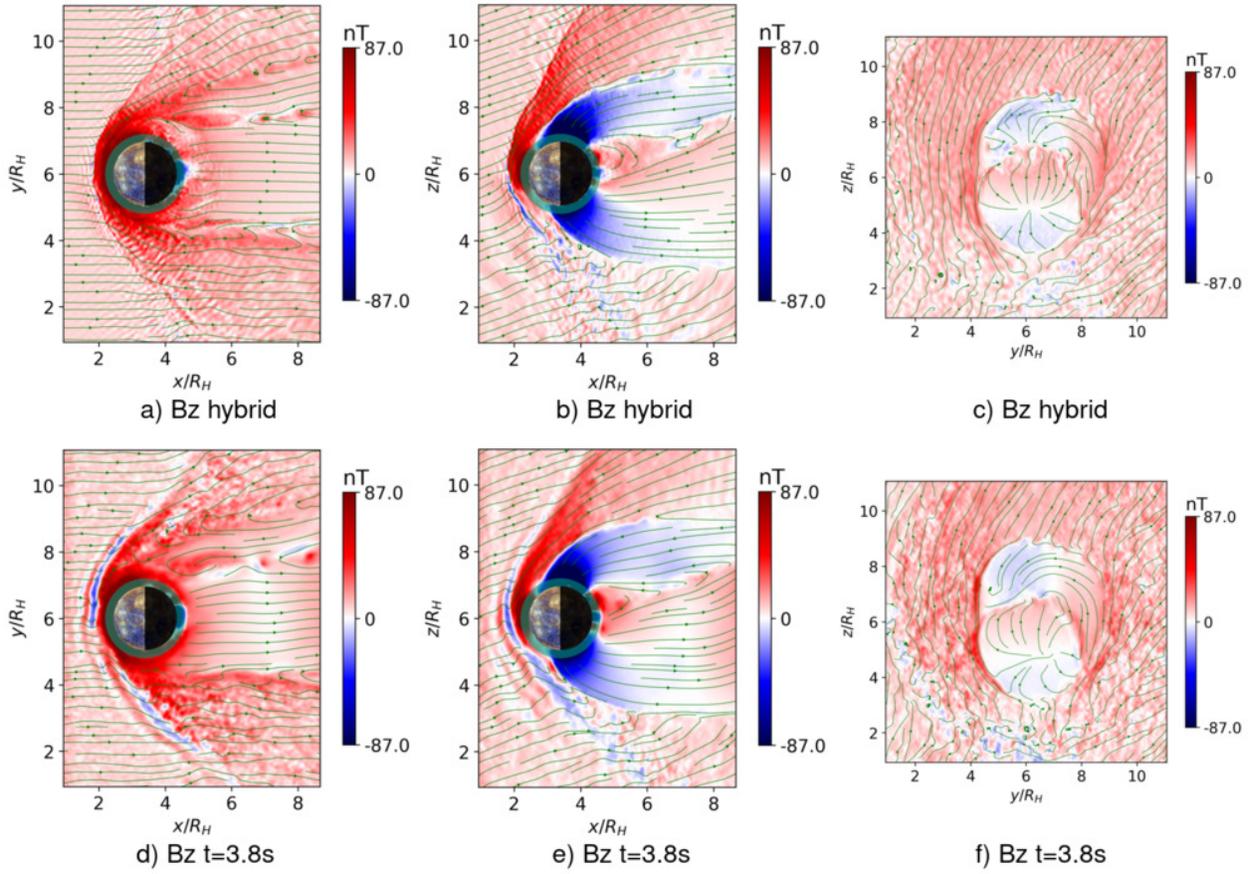

Figure 5: Comparison of the *z* component of the magnetic field, $B_z$ from the hybrid (top) and full kinetic (bottom). The display is the same as in Figure 2.

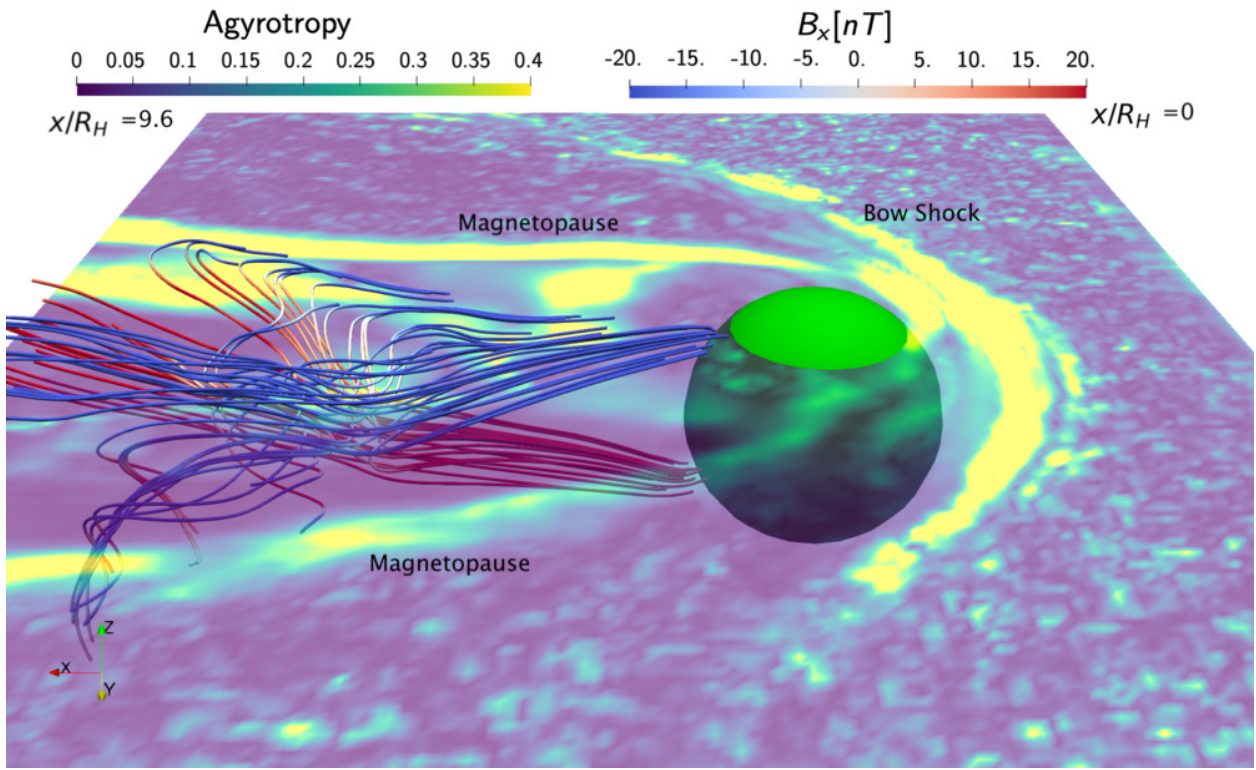

Figure 6: Blow up of the region of tail reconnection. Magnetic field lines undergoing reconnection are superimposed to a false color representation of the electron agyrotropy on the XY plane (at $z/R_H$= 6.8). The filed lines shown are emanating from a region of radius $r/R_H$ =0.5 around $x/R_H$ =6.8, $y/R_H$ =7.2 and $z/R_H$ =6.6.

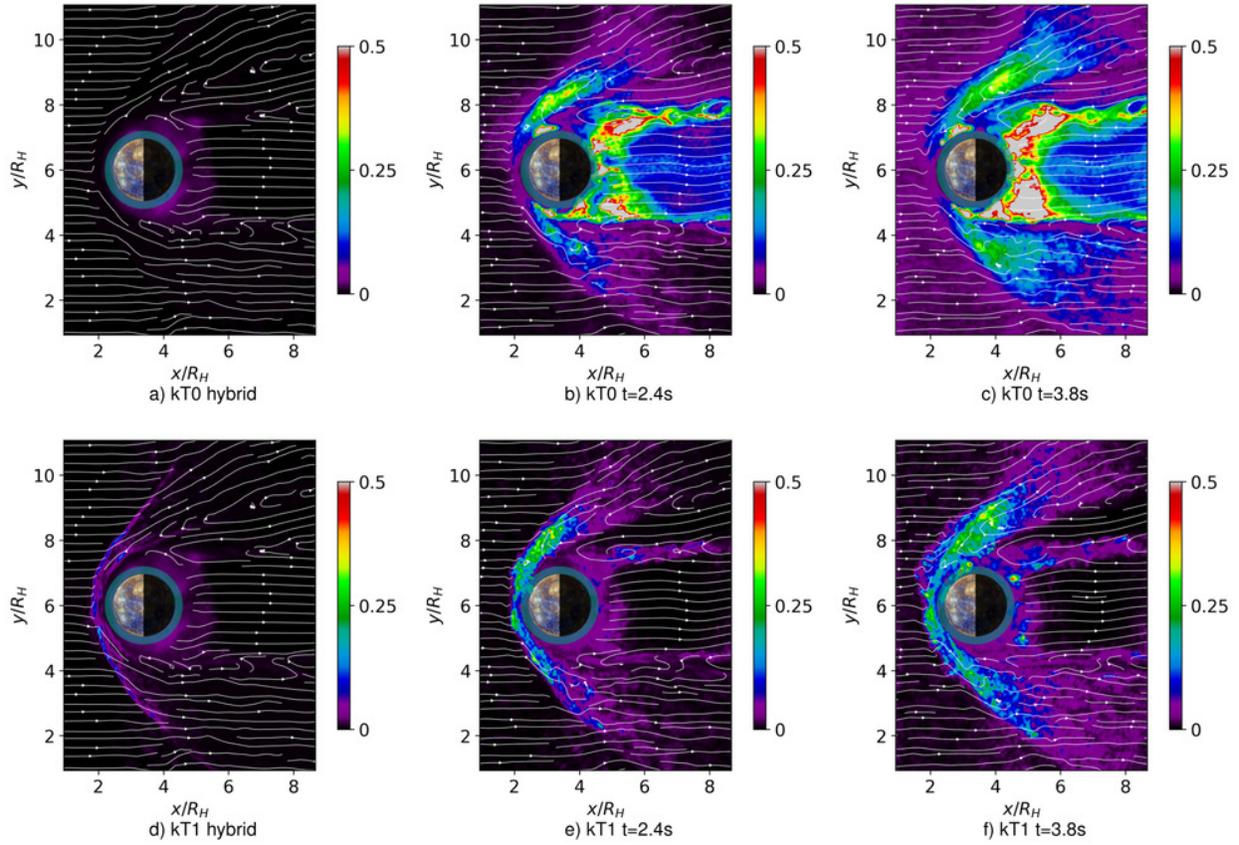

Figure 7: Evolution of the thermal energies, $E_{th,s}$, in the XY plane from the hybrid (top) and 2 subsequent times from the fully kinetic calculation: second row, *Cycle*=100, third row, *Cycle*=1000. The left (right) column shows the electrons (ions). The white lines magnetic field lines projected on the XY plane.

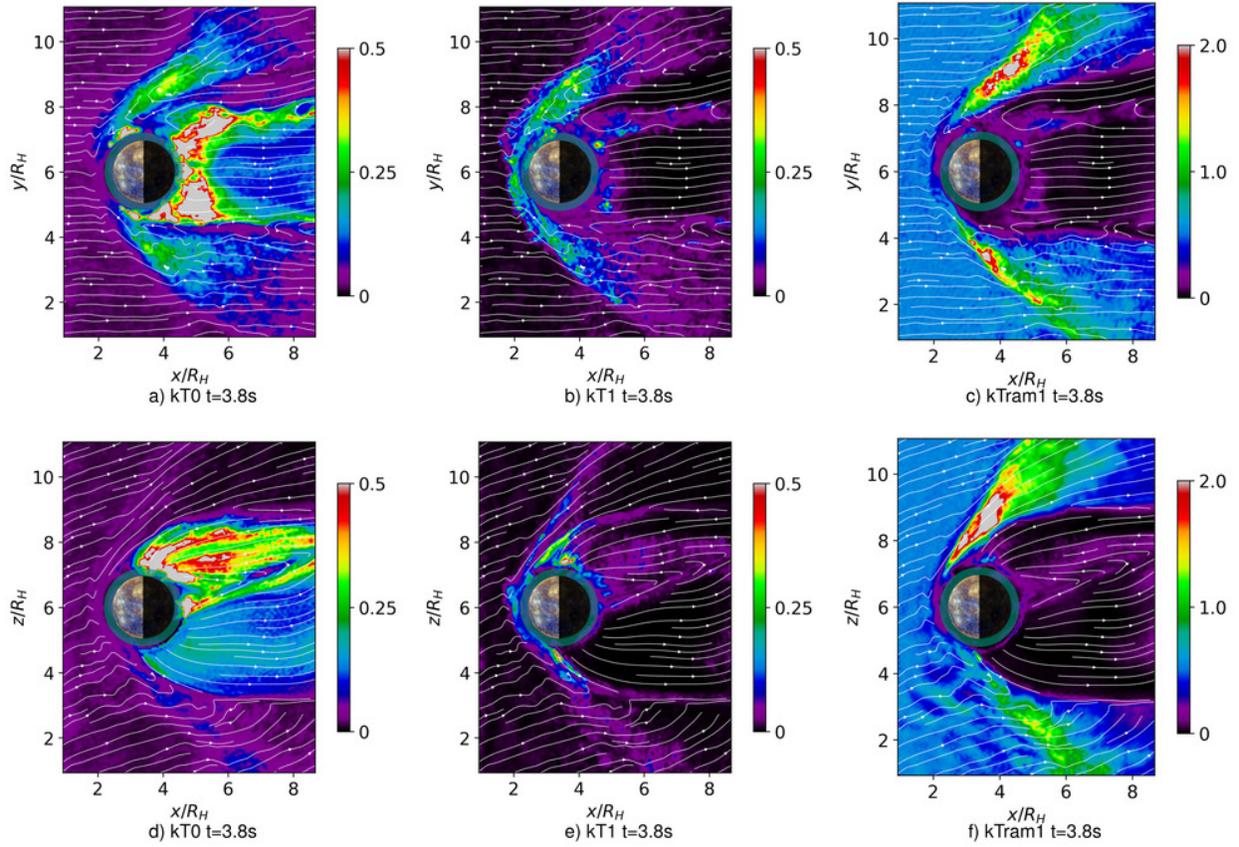

Figure 8: Species energy at the final time (at cycle 1600 or 3.8s after the start) for electrons (left) and ions. The top row shows the *XY* plane and the bottom the *XZ* plane. From left to right shown are $E_{th,e}$, $E_{th,i}$ and $E_i$. The total energy of the electrons is identical and is not reported because the electron bulk energy is negligible. The white lines are projected field lines.

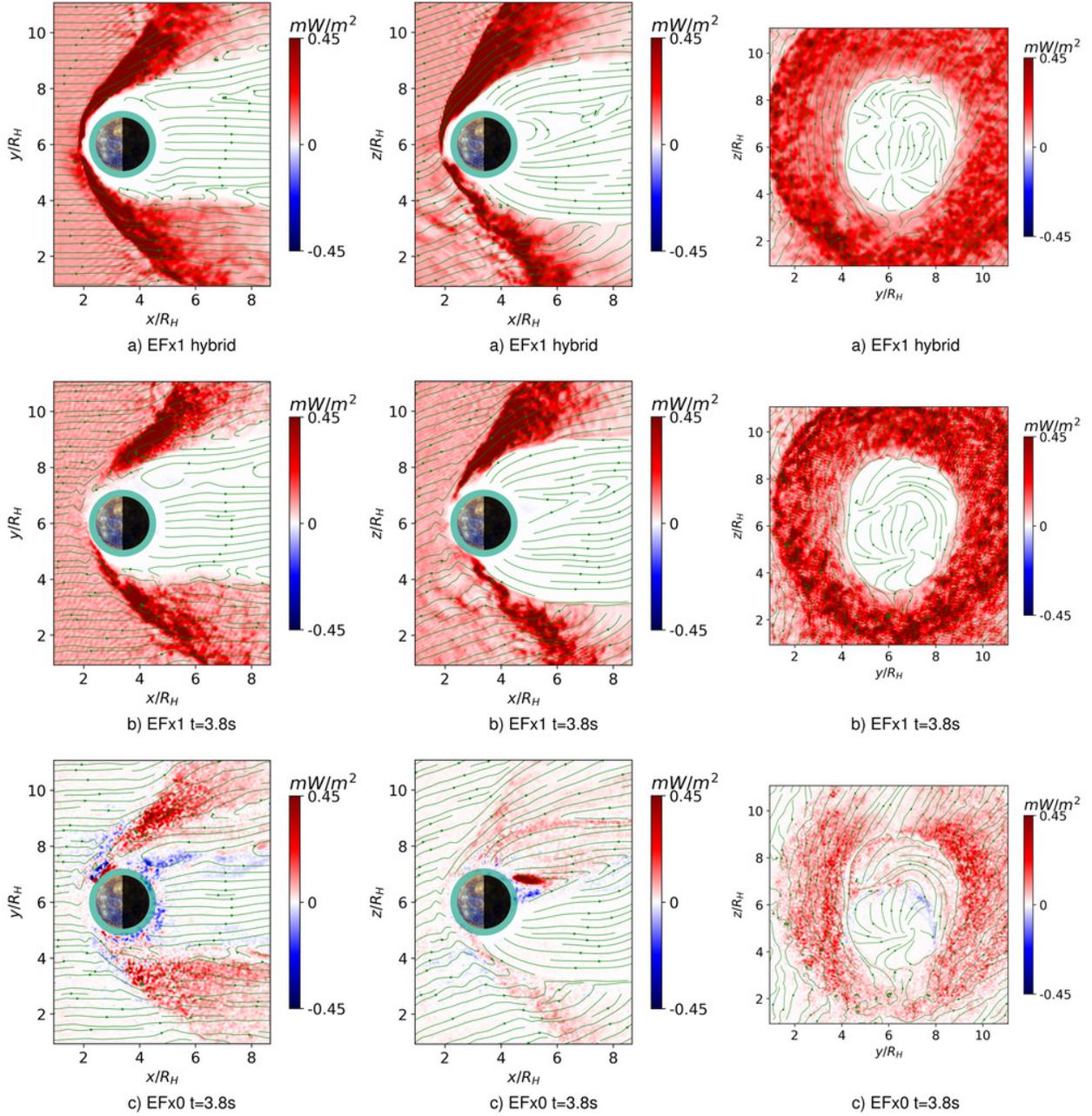

Figure 9: Energy fluxes in the *x* direction $Q_x$. The top row reports the ion energy flow $Q_{x,i}$ from the hybrid run. The next two rows show the final (*Cycle*=1600) energy flow from the full kinetic simulation, for ions $Q_{x,i}$ (middle row) and electrons $Q_{x,e}$ (bottom row).

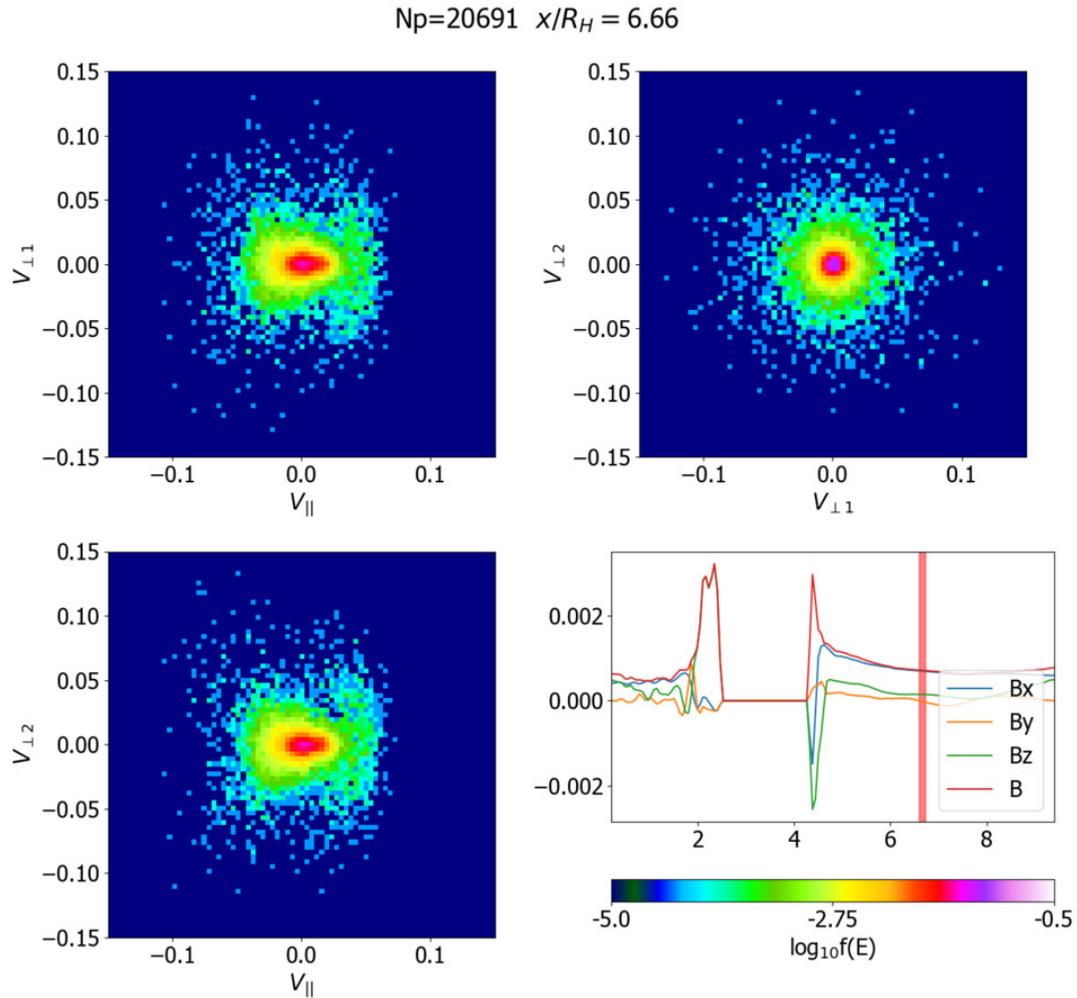

Figure 10: Tail. Reduced electron velocity distribution functions at the location *x* shown in lower right panl by an orange line and $y/R_H=7.5\pm Ly/120$ and $z/R_H =7.5\pm Lz/120$. Three different reduced distributions are shown in magnetic coordinates. Note that compared with Figure 11 the scale of the axes have been expanded to account for the higher thermal speed. The plot of the three components of the magnetic field and the magnitude is along the sun-Mercury line.

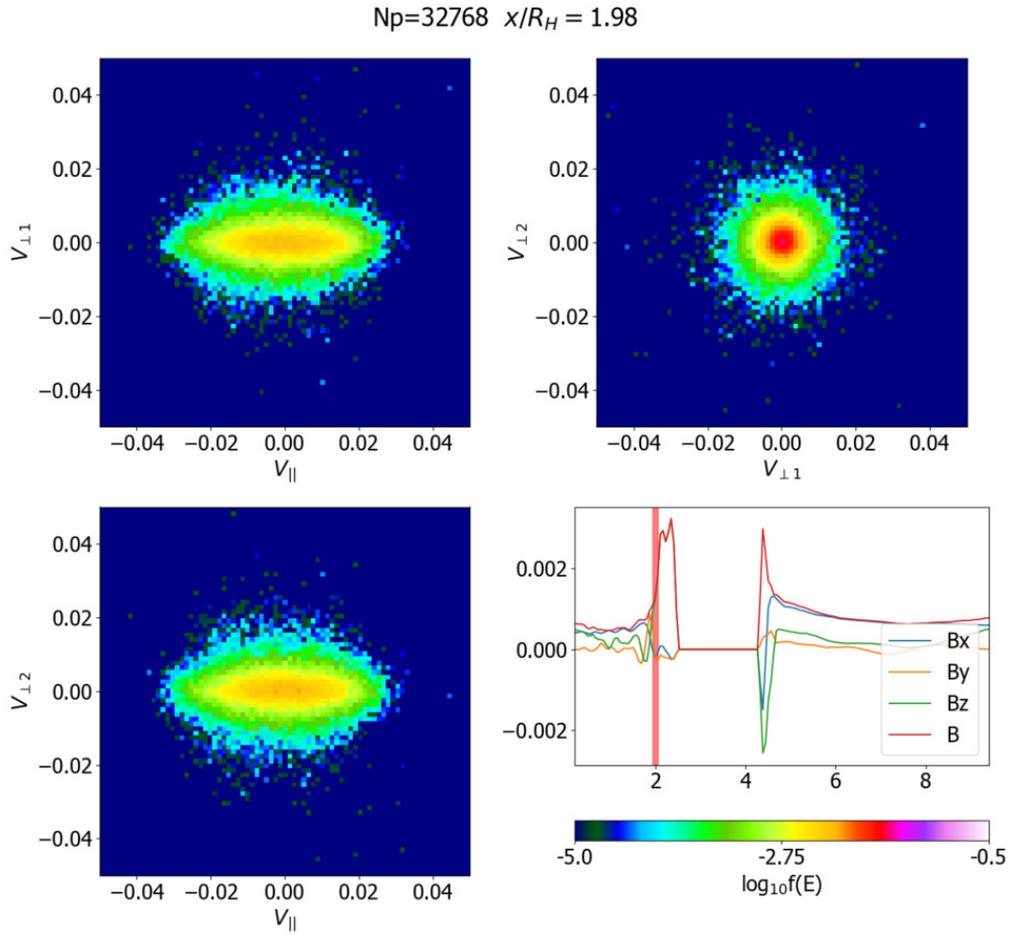

Figure 11: Dayside. Reduced electron velocity distribution functions at the location *x* shown in the lower right panel with an orange line and *y/R$_H$* =7.5±*Ly*/120 and *z/R$_H$* =7.5±*Lz*/120. Three different reduced distributions are shown in magnetic coordinates. The lower right plot shows the three components of the magnetic field and the magnitude along the sun-Mercury line.

Ness, N. F., Behannon, K. W., Lepping, R. P., & Whang, Y. C. (1975). Magnetic field of Mercury confirmed. *Nature*, *255*(5505), 204–205. https://doi.org/10.1038/255204a0

Ness, N. F., Behannon, K. W., Lepping, R. P., & Whang, Y. C. (1976). Observations of Mercury's magnetic field. *Icarus*, *28*(4), 479–488. https://doi.org/10.1016/0019-1035(76)90121-4

Olshevsky, V., Deca, J., Divin, A., Peng, I.B., Markidis, S., Innocenti, M.E., Cazzola, E. and Lapenta, G., 2016. Magnetic null points in kinetic simulations of space plasmas. *The Astrophysical Journal*, *819*(1), p.52.

Omidi, N., Blanco-Cano, X., Russell, C. T., Karimabadi, H., & Acuna, M. (2002). Hybrid simulations of solar wind interaction with magnetized asteroids: General characteristics: SOLAR WIND INTERACTION WITH MAGNETIZED ASTEROIDS. *Journal of Geophysical Research: Space Physics*, *107*(A12), SSH 12-1-SSH 12-10. https://doi.org/10.1029/2002JA009441

Park, J., Lapenta, G., Gonzalez-Herrero, D., & Krall, N. A. (2019). Discovery of an Electron Gyroradius Scale Current Layer: Its Relevance to Magnetic Fusion Energy, Earth's Magnetosphere, and Sunspots. *Frontiers in Astronomy and Space Sciences*, *6*, 74.

Parnell, C. E., Haynes, A. L., & Galsgaard, K. (2008). Recursive reconnection and magnetic skeletons. *The Astrophysical Journal*, *675*(2), 1656.

Russell, C. T., Baker, D. N., & Slavin, J. A. (1988a). The Magnetosphere of Mercury. In *Mercury* (p. 514). University of Arizona Press.

Russell, C. T., Baker, D. N., & Slavin, J. A. (1988b). The Magnetosphere of Mercury. In *Mercury* (p. 514). University of Arizona Press.

Schriver, D., Trávníček, P., Ashour-Abdalla, M., Richard, R. L., Hellinger, P., Slavin, J. A., et al. (2011). Electron transport and precipitation at Mercury during the MESSENGER flybys:
35